\documentclass[aps,amsmath,amssymb,preprintnumbers,a4paper,prd,nofootinbib,twocolumn]{revtex4-1}
 \pdfoutput=1

 \usepackage{lipsum}
\usepackage{bm}
\usepackage{mathrsfs}  

\usepackage{slashed}
\usepackage{subfigure}
\usepackage{feynmp}
\usepackage{graphicx}
\usepackage{amsfonts}
\usepackage{color}
\usepackage{cancel}
\usepackage{amsmath}
\def\hbar{\hspace{0pt}\raisebox{1pt}{$-$} \hspace{-7pt} h}

\def\5{\overline 5}

\definecolor{JJ}{RGB}{0,144,255}

\newcommand{\be}{\begin{equation}}
\newcommand{\ee}{\end{equation}}
\newcommand{\bea}{\begin{eqnarray}}
\newcommand{\eea}{\end{eqnarray}}

\newcommand{\ba}{\begin{eqnarray}}
\newcommand{\ea}{\end{eqnarray}}

\begin{document}
%%%%%%%%%%%%%%%%%%%%%%%%%%%%%%%%%%%%%%%%%%%%%%%%%%%%%%%%%%%  FRONT PAGE
\title{{Minimal Composite Dynamics versus Axion Origin  of the Diphoton excess}}

\author{Emiliano Molinaro}

\author{Francesco Sannino}

\author{Natascia Vignaroli}
 \affiliation{CP$^{3}$-Origins and the Danish IAS, University of Southern Denmark, Campusvej 55, DK-5230 Odense M, Denmark}

\begin{abstract} 

ATLAS and CMS observe deviations from the expected background  in the diphoton invariant mass searches of new resonances around 750 GeV. We show that  a simple realization in terms of a new pseudoscalar state can accommodate the observations. The model leads to further footprints that can be soon observed.  The new state can be interpreted both as an axion or as a {highly natural} composite state arising from minimal models of dynamical electroweak symmetry breaking. We further show how to disentangle the two scenarios. Beyond the possible explanation of the diphoton excess the results show that it is possible to directly test and constrain composite dynamics via processes stemming from its distinctive topological sector. 

\preprint{CP3-Origins-2015-052 DNRF90, DIAS-2015-52}
 
\end{abstract}
 %%%%%%%%%%%%%%%%%%%%%%%%%%%%%%%%%%%%%%%%%%%%%%%%%%%%%%%%%%%%%%%%%%%

\maketitle

 The ATLAS  and CMS \cite{ATLAS,CMS} find local excesses respectively of 3.6$\sigma$ and 2.6$\sigma$ for a resonance in diphoton searches with invariant mass spectrum around 750 GeV. This leads to  a global significance of 2$\sigma$ in ATLAS and 1.2$\sigma$ in CMS. 

The ATLAS and CMS results in the $a \rightarrow 2 \gamma$  channel suggest a reconstructed mass of around $750$~GeV  and a cross-section $\sigma (pp \to a\to 2\gamma)$ of the order of $6 ~{\rm fb} $. Here $a$ denotes a new intermediate massive state.

We assume $a$ to be a neutral spin zero particle and employ a minimal description in terms of an effective field theory,  and study the resulting phenomenology in the narrow width approximation.  After introducing the relevant effective operators  we show how they can emerge within two calculable extensions of the Standard Model (SM) featuring either a new elementary axion-like or a composite $\eta^\prime$-like state. We further assume both states to couple to a new colored vectorlike fermion $T$. The underlying realisations allow us to make further predictions and relate some of the effective couplings.  The models encompass all the needed ingredients to describe the {\it signal channels} and relevant constraints. We  will also provide distinctive signatures aimed at disentangling the composite nature from the elementary one.

We then make contact with time-honoured models of minimal composite dynamics \cite{Weinberg:1975gm,Susskind:1978ms,Kaplan:1983fs,Kaplan:1983sm}. Weinberg and Susskind's minimal models of dynamical electroweak (EW) symmetry breaking \cite{Weinberg:1975gm,Susskind:1978ms}  are based on QCD-like dynamics and  are at odds with experiments. {Within this early model realisations one finds the pioneering work of Di Vecchia and Veneziano \cite{DiVecchia:1980xq} that long ago envisioned a scenario similar to the one presented here.} Modern incarnations that are still  minimal but employ non-QCD like dynamics are phenomenologically viable \cite{Sannino:2004qp,Dietrich:2005jn,Ryttov:2008xe}.  Complementary signal channels for spin-one resonances have  been investigated in more complete model implementations, e.g. in \cite{Belyaev:2008yj}, that can even explain the  2-TeV diboson excess \cite{Franzosi:2015zra}.  
The general features, regarding resonance mass, cross-section and decay patterns are very much in line with models of weak scale compositeness \cite{Sannino:2004qp,Dietrich:2005jn,Ryttov:2008xe,Belyaev:2008yj}. 

Therefore the minimal weak-scale composite paradigm, besides solving the hierarchy problem:  
\begin{itemize} 
\item{explains the diphoton excess,}
\item{ naturally accommodates the $750$~GeV mass,}
\item{predicts and relates new processes,}
\item{can be disentangled from other less natural models.}
\end{itemize}
Amusingly, the best fit value for the decay constant of the new state is highly compatible with the one needed to break the EW symmetry dynamically. 

%\subsection*{Effective operators} 
 Assuming the existence of a new pseudoscalar state $a$  the CP conserving effective operators linking the spin-zero resonance with the SM fermions are
\bea
\mathcal{L}^{a}_{\rm qq}&= - \sum_{ij}   i\,a \,\bar{q}_i  \, Y^a_{ij}\, \gamma_5  q_j \ ,\label{a-q-q}
\eea
where $i,j$ runs over all  flavors.  The effective operators linking our states to SM gauge bosons are
\bea
\mathcal{L}^{a}_{V\widetilde{V}} = - \sum_{V_1 \widetilde{V}_2}  \frac{ g_{V_1 \widetilde{V}_2}}{8} \, a\, V_1 \widetilde{V}_2 \ ,
\label{eq:gauge}
\eea
with
 \bea
\sum_{V_1 \widetilde{V}_2}  {g}_{V_1 \widetilde{V}_2}   \, V_1 \widetilde{V}_2  & = &  {g}_{G\widetilde{G}} \,{\rm Tr}[G_{\mu \nu} \widetilde{G}^{\mu \nu}] + {g}_{A  \widetilde{A}}A_{\mu \nu} \widetilde{A}^{\mu \nu} 
 \nonumber \\ 
& + &g_{Z \widetilde{Z}}Z_{\mu \nu} \widetilde{Z}^{\mu \nu}   + 2  {g}_{W {\widetilde{W}}} W^+_{\mu \nu} {{\widetilde{W}}}^{-\mu \nu} 
\nonumber \\ &+&   2 g_{Z \widetilde{A}}Z_{\mu \nu}{\widetilde{A}}^{\mu \nu}  
\eea
and  $ \widetilde{V} ^{\mu\nu}= \epsilon^{\mu\nu\rho\sigma}V_{\rho \sigma}$.  
  Depending on the underlying realisation some of these couplings might either vanish, develop hierarchies and/or be related to each other.  In the following we will consider the case in which
  the effective couplings in (\ref{a-q-q}) are negligible.
  
\subsection*{Axion Realisation}
  We add  to the SM the terms: 
 \bea
 \Delta{\cal L}  & =&\frac{1}{2}\left( \partial_\mu a \,\partial^\mu a - m_a^2 a^2  \right)  + i\overline{T} \gamma_\mu D^\mu T -  i y_T \, \frac{m_T}{f_a} \,  a \, \overline{T} \gamma_5 T \nonumber \\ 
 & - & m_T \overline{T}T  +  \Delta_{t-T}^{\rm mix}\label{ytcoupling}
 \eea
  where $T$ is an SU(2) weak singlet vectorlike quark in a given representation of color interactions. The new Yukawa interaction strength is controlled by $y_T$ and the $a$ decay constant $f_a$. The mixing mass-term operator 
$\Delta_{t-T}^{\rm mix}$   between the top and the new colored state vanishes unless the representation of  $T$ is the fundamental of color and/or the hypercharge matches. The action above can represent that of an axion-like state. We will see later that a more natural interpretation emerges when this state is viewed as a composite one, provided new operators are added stemming from its topological sector.  
  
  $T$ loops, in the fundamental representation of color, generate the following effective couplings 
\begin{eqnarray}
%\begin{split}
\label{eq:effective-couplig}
 g^T_{G\widetilde{G}} & = &y_T \frac{\alpha_S}{2\,\pi\, f_a} F\left(\frac{m_a^2}{4\,m_T^2}\right), ~
 g^T_{A\widetilde{A}} =  y_T \frac 43 \frac{\alpha_\text{em}}{\pi\, f_a} F\left(\frac{m_a^2}{4\,m_T^2}\right)\, , \nonumber \\
	g^T_{A\widetilde{Z}} & = & \tan(\theta_W)\, g_{A\widetilde{A}}\ , \quad g^T_{Z\widetilde{Z}} \ = \  \tan^2(\theta_W)\, g_{A\widetilde{A}}\,,
%\end{split}
\end{eqnarray}
where $\theta_W$ is the weak mixing angle and, in the approximation $m_a<2\,m_T$, we have 
 	$F(x) = \arcsin\left(\sqrt{x}\right)/x  \approx  1\,+\,x/ 3\,+\, 8\,x^2/45 \,+\,4\,x^3/35 $.
 From the model we arrive at the following relevant partial decay rates
 \begin{eqnarray}
 	\Gamma(a \to g g) &= & \frac{m_a^3}{8\,\pi}\, \left(g_{G\widetilde{G}}^{T}\right)^2\,,\\ 
	\Gamma(a \to \gamma\gamma) &= & \frac{m_a^3}{64\,\pi}\, \left(g_{A\widetilde{A}}^{T}\right)^2\,,\\
	\Gamma(a \to \gamma Z) &= & \frac{m_a^3}{32\,\pi}\,\left(g_{A\widetilde{Z}}^{T}\right)^2\left(1\,-\,\frac{m_Z^2}{m_a^2}\right)^3, \\
	\Gamma(a \to Z Z) &= & \frac{m_a^3}{64\,\pi}\,\left(g_{Z\widetilde{Z}}^{T}\right)^2\left(1\,-\,\frac{4\,m_Z^2}{m_a^2}\right)^{3/2} ,
 \end{eqnarray}
 and branching ratios 
 \begin{eqnarray}\label{BRaxion}
 	B(a\to\gamma\gamma) &\approx& 0.0063\,,\nonumber\\
	B(a\to \gamma Z) &\approx& 0.0037\,,\nonumber\\
	B(a\to Z Z) &\approx& 0.00055\ . 
 \end{eqnarray}
 From the above it is clear that the branching ratio into two photons dominates with respect to the other EW channels. Furthermore, while the partial widths depend directly on $f_a/y_T$ and $m_a$ the branching ratios depend only on  $\alpha_S$ and weak coupling constants.

We implement our model in MadGraph5\_aMC@NLO \cite{Alwall:2011uj} with the effective couplings in (\ref{eq:effective-couplig}) and we calculate the $a$ production cross-section at leading order. The QCD next-to-leading order
(NLO) corrections are calculated %in MadGraph5\_~aMC at NLO by 
using the model provided in \cite{Demartin:2014fia}. For $m_a \simeq 750$ GeV we obtain a NLO $K$-factor: $K^{NLO}\simeq 2.6$. 
 
 \begin{figure}
 \includegraphics[width=0.45\textwidth]{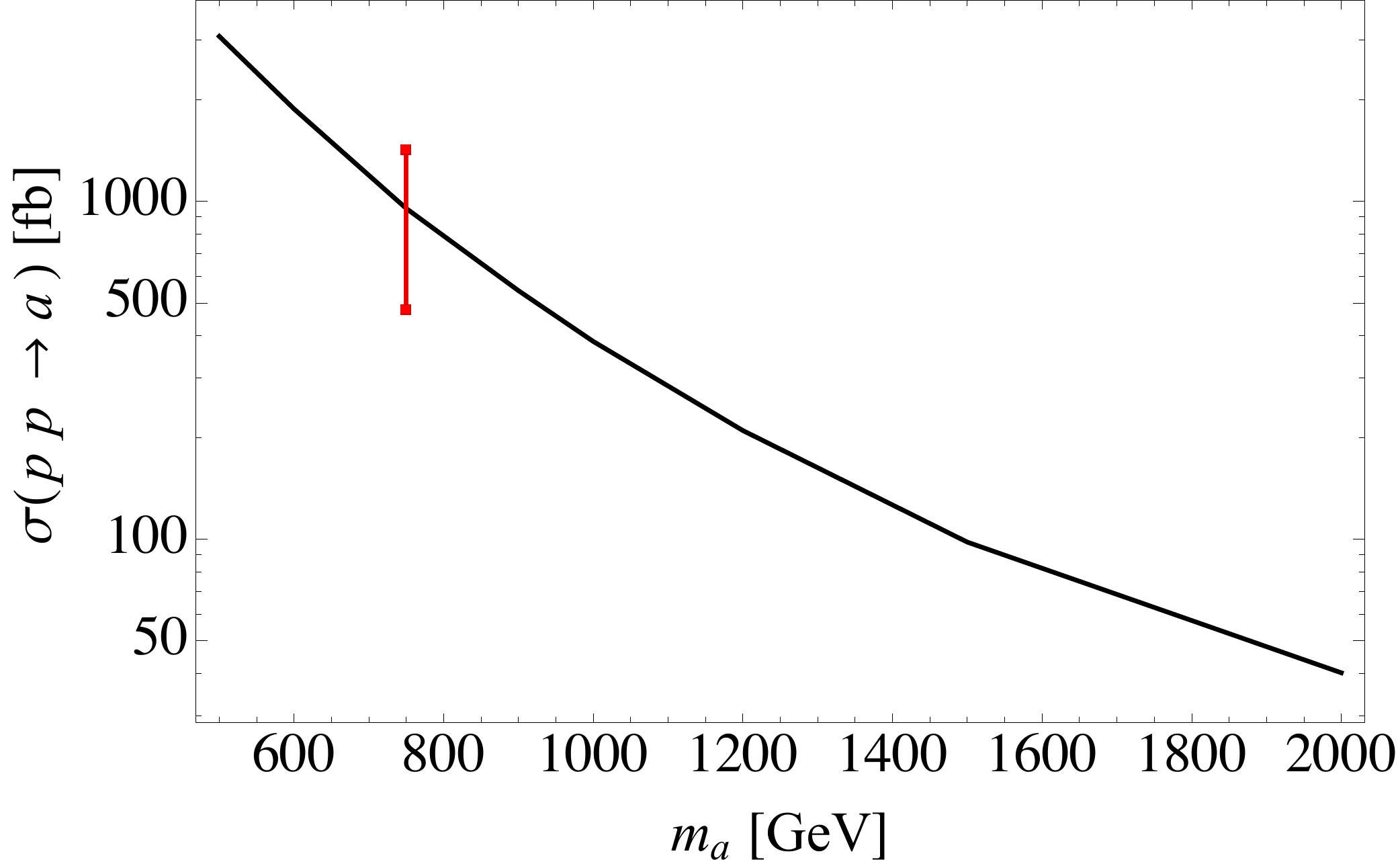}
 \caption{Production cross-section  of  $a$ at the 13 TeV LHC as function of the pseudoscalar mass, for the value $f_a/y_T=245$ GeV reproducing the ATLAS excess and $m_T=1$ TeV. The vertical bar at 750 GeV shows the $\pm 1\sigma$ deviation from the central value of the excess. The cross-section scales as $y^2_T/f^2_{a}$. }
 \label{fig:xsec}
 \end{figure}
 
 Fig.~\ref{fig:xsec} shows the production cross-section of the pseudoscalar $a$ at the LHC with $\sqrt{s}=13$ TeV as function of $m_a$ for the specific $f_a/y_T$ value which reproduces the ATLAS excess, and for a 1~TeV  vectorlike quark $T$.  
 
 A fit to the ATLAS excess, in the narrow-width-approximation, gives the value  $\sigma (pp \to a\to 2\gamma)=(6 \pm 3)$~fb. Remarkably this value is reproduced by our model for an $f_a$ scale
 \begin{equation}\label{eq:f}
 f_a/|y_T|=245^{+100}_{-45} \ \text{GeV} \ ,
 \end{equation}
    with $m_T=1$ TeV. We find a very mild dependence of our results on the $T$ mass. 
  A lower  limit on $m_T$ of the order of $800$ GeV is placed by the run-1 LHC results \cite{CMS:2014dka, ATLAS:2015dka}. The vectorlike quark coupling to $a$, $y_T m_T/f_a$, reaches the non-perturbative regime for $m_T \gtrsim 1/y_T$ TeV. \\
  On top of the presence of the vectorlike $T$, possibly within the reach of the current LHC run \cite{Vignaroli:2012nf}, the model predicts a significant $a$ decay branching ratio to $Z\gamma$. The corresponding signal with $Z$ decaying to leptons could be observed when the experiments will have collected roughly $100$ fb$^{-1}$. This would  correspond to a number of expected signal events comparable with those in the $\gamma\gamma$ excess. The narrow $a$ resonance can also be detected via di-jets. However, the sensitivity of the 8 TeV searches \cite{Aad:2014aqa,CMS:2015neg} in this channel is low, due to the overwhelming QCD background at the relatively low di-jet mass spectrum near 750 GeV.  In fact, the CMS search \cite{CMS:2015neg} places an upper limit on the di-jet cross-section times acceptance of about 1.8 pb for a resonance at 750 GeV, which is an order of magnitude above the value we obtain for the $a$ pseudoscalar, that is $\sim$70 fb, assuming an acceptance of 0.6.  {The run-1 LHC searches in the $\gamma\gamma$ channel at $\sqrt{s}=8$ TeV \cite{Aad:2015mna,Khachatryan:2015qba} are also compatible with our explanation of the diphoton excess at 13 TeV. The 95\% C.L. upper limit on the $\gamma \gamma$ cross-section at the invariant mass $m_{\gamma\gamma}\sim 750$ GeV is of about 1.5 fb, while the axion model predicts about 0.7 fb.} 
  The present explanation of the ATLAS and CMS diphoton excesses, therefore, passes all of the current experimental tests.
   
 \subsection*{Minimal Composite Dynamics}
  
We move now to the interpretation of the new state within minimal (near-conformal) models of dynamical EW symmetry breaking \cite{Appelquist:1998xf,Appelquist:1999dq,Duan:2000dy,Sannino:2004qp,Dietrich:2005jn,Dietrich:2005wk,Dietrich:2006cm,Foadi:2007ue}. 

In particular the new  heavy resonance can be identified, in composite models at the EW scale, with the singlet $\eta^\prime$ state that we rename here $a$. This requires a new strongly coupled sector featuring $N^F_{T}$ Dirac flavors transforming under the representation $R$ of a new gauge sector $SU(N_{T})$ with $N_T$ the new color.

Assuming that the number of flavors is barely outside the conformal window \cite{Dietrich:2006cm}, the new gauge sector will 
generate chiral symmetry breaking with confining scale $\Lambda_T$ and technipion decay constant $F_T$. 
At energies below $\Lambda_{T}$ the relevant degrees of freedom, needed for the present analysis, are encoded in the unitary matrix
\begin{equation}
	U\; = \; e^{i \Phi /F_T}\;=\; \exp\left[ \frac{i}{F_T}\left(a\,+\, {\vec{\tau}}\cdot {\vec{\pi}}\right)  \right]\,,
\end{equation}
where $\vec{\tau}\equiv\left(\tau_1,\,\tau_2\,,\tau_3 \right)$ are the standard Pauli matrices, whereas $a$ 
and $\vec{\pi}\equiv\left(\pi^ 1,\,\pi^2\,\,\pi^3\right)$ are, respectively, the singlet and isotriplet pseudoscalar resonances. For simplicity we assume $N^F_T=2$, but it is straightforward to generalise this to a different number of flavors. The associated quantum global symmetry is  $SU(2)_L \times SU(2)_R$. Although a larger symmetry group can be considered for a possible realisation within composite Goldstone Higgs dynamics  we focus here on minimal composite realisations.\footnote{Here  the underlying  dynamics does not carry color. This guarantees minimality, meaning that no new colored technihadron states are present in the spectrum. Uncolored technifermions are also preferred by precision EW observables.}~Of course, we should also include a Higgs-like state. Depending on the explicit underlying dynamics, it might emerge either as the lightest dilaton-like state stemming from a near conformal theory  \cite{Dietrich:2005jn,Foadi:2012bb} or, if the symmetry is opportunely enhanced, as pseudo Goldstone Boson \cite{Kaplan:1983fs,Kaplan:1983sm}.  Furthermore near-conformality\footnote{ Analytical \cite{Sannino:2004qp,Dietrich:2006cm,Bergner:2015dya} and numerical efforts \cite{Catterall:2007yx,Hietanen:2009zz,DelDebbio:2010hx,DeGrand:2011qd,Appelquist:2011dp,DeGrand:2010na,Fodor:2015zna,Hasenfratz:2015ssa,Athenodorou:2014eua} have been dedicated to determine whether fermionic gauge theories display large distance conformality. For the sextet model lattice results suggest that the theory is either very near-conformal or conformal. In the latter case  interactions responsible for giving masses to the SM fermions can modify the conformal-boundary inducing an ideal near-conformal  behaviour \cite{Fukano:2010yv}.}  alleviates tension with EW precision measurements \cite{Appelquist:1998xf} and flavor changing neutral currents constraints \cite{Holdom:1983kw}.

 The effective action generating the relevant $a$ interaction is: 
\begin{equation}\label{effaction}
\Gamma = \int d^4x\left( \mathcal{L}_0\,+\,\mathcal{L}_{m_{a}} \right) \,+\,\Gamma_{WZW}\,.
\end{equation}
Each term in (\ref{effaction}) can be expressed in terms of $U$, the Maurer-Cartan one-forms
\begin{eqnarray}
\alpha & = & \left(\partial_{\mu}U\right)U^{-1} dx^{\mu}\equiv
\left(dU\right)U^{-1},~~
 \beta  =  %U^{-1}dU=
 U^{-1}\alpha U  \label{MC}
\end{eqnarray}
and additional ``left" and ``right" one-forms, $A_L=A^\mu_L dx_\mu$ and $A_R=A_R^\mu dx_\mu$, respectively, with
\begin{eqnarray}
 A^\mu_L &=&   g_Y \left(Q-\frac 12 \tau_3 \right)\,B_\mu\,+\, \frac 12 g_W\, \vec{\tau}\cdot\vec{W}^\mu\,, ~~
  A^\mu_R =  g_Y Q  B^\mu\,, \nonumber
\end{eqnarray}
where $Q$ denotes the electric charge matrix of the fundamental technifermions.

The relevant Lagrangian terms are\begin{eqnarray}
	\mathcal{L}_0 & = & \mathcal{L}_{\rm kin} 
	                       +  \frac{F_T^2}{2}\,{\rm Tr}\left[ \partial_\mu \xi^\dagger \partial ^\mu\xi \right]-
	                       \frac{F_T^2}{2}\,{\rm Tr}\left[ \xi^\dagger\partial_\mu\xi \partial^\mu\xi\xi^\dagger\right]\nonumber\\
	                       &+& \frac{F_T^2}{4}\,{\rm Tr}\left[ A_L^\mu A_{L\mu}+ A_R^\mu A_{R\mu}  \right]
	                       - \frac{F_T^2}{2}\,{\rm Tr}\left[ A_L^\mu U A_{R\mu} U^\dagger\right] \nonumber \\
	                       &-& \frac{i\,F_T^2}{2}\,{\rm Tr}\left[ \partial_\mu \xi\, \xi^\dagger A_L^\mu -\xi^\dagger\, \partial_\mu \xi A_R^\mu\right]\nonumber\\
	                       &-& \frac{i\,F_T^2}{2}\,{\rm Tr}\left[\xi\,\partial_\mu\xi \,U^\dagger \,A^\mu_L - \partial_\mu\xi\,\xi^\dagger \,A_R^\mu\,U^\dagger \right]\,,
\end{eqnarray}
where $\xi\equiv U^{1/2}$ and $\mathcal{L}_{\rm kin}$ is the standard kinetic term for the vector fields. The second term in (\ref{effaction}) comes from a quantum anomaly and it provides 
a mass term for the singlet $a$.  It reads
\begin{equation}
	\mathcal{L}_{m_{a}} \;= \; \frac{\kappa F_T^2}{8\,N_T}{\rm Tr}\left[ \ln U -\ln U^\dagger\right]^2\,,
\end{equation}
 and $\kappa$ is connected to the mass of $\eta_0$ 
($m_{\eta_0}=849$~MeV),~\footnote{{Here $\eta_0$ indicates the QCD SU(3) flavor singlet state in the chiral limit, with \cite{Veneziano:1979ec} 
\begin{equation}
	m_{\eta_0}^2\;=\; m_{\eta^{\prime}}^2\,+\,m_\eta^2\,-\,2\,m_K^2\,.
\end{equation}}}
\begin{equation}
	\kappa\;=\;\frac16\frac{F_T^2}{f_\pi^2}\frac{9}{N_T}\,m_{\eta_0}^2\,.
\end{equation} 
Then, the mass of $a$ (taking $f_\pi=92$ MeV), for technifermions in the fundamental representation, is 
\begin{equation}\label{mamass}
	m_a \; = \; \sqrt{\frac{2}{3}}\frac{F_T}{f_\pi}\,\frac{3}{N_T}\,m_{\eta_0}\,\approx\, \frac{6}{N_T}~\text{TeV}\,.
\end{equation}
 Thus, we have that a 750 GeV $a$-state naturally emerges in this scenario for $N_T=6$ and $8$.
 Similar values of $m_a$ could emerge from near conformal field theories with smaller $N_T$  \cite{Sannino:2004qp,DiVecchia:2013swa}.

As for the relevant terms arising from the gauged Wess-Zumino-Witten operator \cite{Witten:1979vv,Veneziano:1979ec,DiVecchia:1980xq,Kaymakcalan:1983qq} we have:
\begin{eqnarray}
&& \Gamma _{WZW}\left[ U,\;A_{L},\;A_{R}\right] =\Gamma _{WZ}\left[ U\right]
\,
\nonumber \\
&&-5C\,\int_{M^{4}}{\rm Tr}\left[ (dA_{L}A_{L}+A_{L}dA_{L})\alpha
+(dA_{R}A_{R}+A_{R}dA_{R})\beta \right]  \nonumber \\
&&+5C\,\int_{M^{4}}{\rm Tr}\left[
dA_{L}dUA_{R}U^{-1}-dA_{R}dU^{-1}A_{L}U \right] 
  + \cdots  ,  \label{GWZ1}
\end{eqnarray}
where $C=-i d(R)/(240\,\pi^2)$, $d(R)$ is the dimension of the technifermion representation. For the fundamental representation $d({\rm Fund}) = N_T$. 
Here  $F_{L}$ and $F_{R}$ are  two-forms defined as
$F_{L}=dA_{L}-iA_{L}^{2}$ and $F_{R}=dA_{R}-iA_{R}^{2}$. 
The Wess-Zumino effective action is $\Gamma_{WZ}\left[U\right]=C\, \int_{M^5} {\rm Tr} \left[\alpha^5\right]$.
From (\ref{GWZ1}) we have extra contributions to the effective couplings to the gauge bosons given in (\ref{eq:gauge}) that alter the predictions with respect to the axion-like scenario. These extra contributions are:  
\begin{eqnarray}
	g^{\rm comp}_{A\widetilde{A}} &=& \left(1+y^2\right)\frac{e^2}{F_T}\frac{d(R)}{8\,\pi^2}\,,
\end{eqnarray}
\begin{eqnarray}
	g^{\rm comp}_{A\widetilde{Z}} &=& \frac{1-2(1+y^2)s_W^2}{2\,c_W\,s_W}\frac{e^2}{F_T}\frac{d(R)}{8\,\pi^2}\,,
\end{eqnarray}	
\begin{eqnarray}	
	g^{\rm comp}_{Z\widetilde{Z}} &=& \frac{e^2}{F_T} \frac{1-3s_W^2+3(1+y^2)s_W^4}{3\,c_W^2\,s_W^2}\frac{d(R)}{8\,\pi^2}\,,
\end{eqnarray}	            
\begin{eqnarray}	            
	 g^{\rm comp}_{W\widetilde{W}} &=& \frac{e^2}{F_T}\frac{1}{s_W^2}\frac{d(R)}{24\,\pi^2}\,,           
\end{eqnarray}
 $y$ being the hypercharge of the (left-handed) fundamental technifermions \cite{DiVecchia:1980xq} with the normalisation  $Y(Q_{L}) = y /2$ and   $Y(U_{R}/D_{R}) = (y \pm 1)/2$ and  $Q = Y + \tau_3/2$. Here $Q_L$ is the techniquark $SU(2)_L$ transforming doublet and $Q_R=(U_R, D_R)$ the singlets\footnote{The possible presence of the Witten \cite{Witten:1982fp} and/or SM gauge anomalies depends on the underlying technifermion representation  and hypercharge assignment. If present, one can cancel them by adding new lepton-like heavy fermions  \cite{DiVecchia:1980xq,Dietrich:2005jn}.  Direct limits on stable singly-charged SM-like heavy fermions require them to be heavier than about $574$~GeV \cite{Chatrchyan:2013oca}.  
Since the couplings of the new leptons to SM degrees of freedom depend on unknown dynamics, these limits should be taken {\it cum grano salis}.}.

 To compare with the elementary scenario, we added the same coupling in (\ref{ytcoupling}) between the composite $a$ and the new fermion $T$. It could be generated via instantons  from an extended gauge dynamics (EGD). Near the EW scale the EGD would lead to the effective 6-fermion operator ${\rm det}\left( \bar{Q}_L  Q_R \right)\bar{T}_L T_R  + {\rm h.c.}$  suppressed by  $\Lambda_{\rm EGD}^5$. Since
the  UV scale $\Lambda_{\rm EGD}$ is expected to be larger than the EW scale, this operator would typically lead to small values of $y_T$.

The effective couplings of $a$ to the EW gauge bosons are therefore given by $g_{V\widetilde V}=g^{comp}_{V \widetilde V}+g^{T}_{V \widetilde V}$,
where $g^{T}_{V \widetilde V}$ are reported in (\ref{eq:effective-couplig}).
We can now calculate the region of the parameter space which allows to reproduce the diphoton excess in the $\eta^\prime$-like scenario introduced above.
We assume $F_T=246$ GeV and $d({\rm Fund}) = N_T=6$ in such a way to naturally reproduce the pseudoscalar mass $m_a\approx 750$ GeV, see eq.~(\ref{mamass}).
We can also have $d(R)=6$ when the technifermions are in a symmetric of $SU(N_T)$ with $N_T=3$.  {Precision observables were studied in  \cite{Dietrich:2005jn}.}

\begin{figure*}[t!]
\begin{center}
\includegraphics[width=0.36\textwidth]{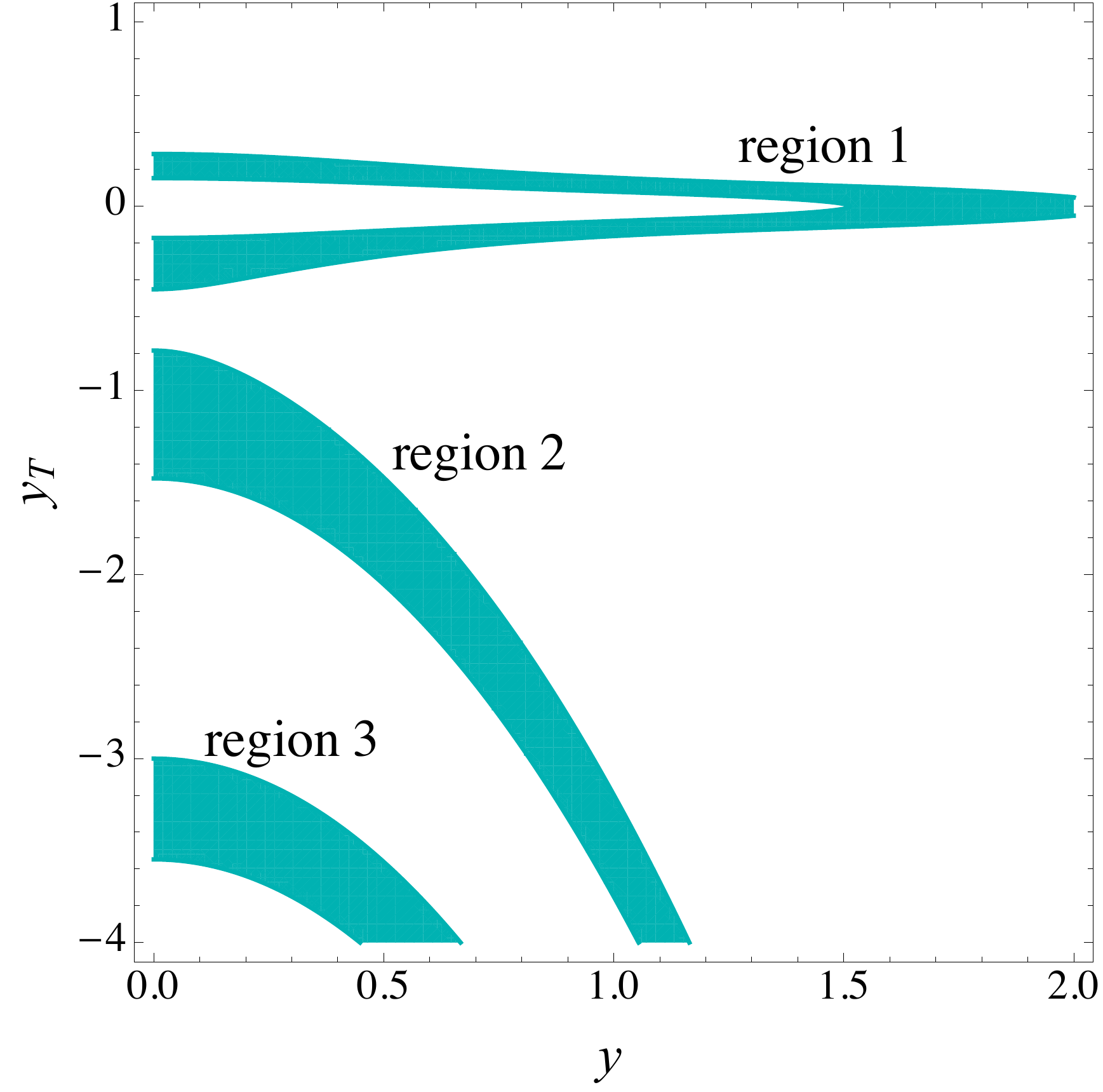} \,~~~~~~~~
\includegraphics[width=0.36\textwidth]{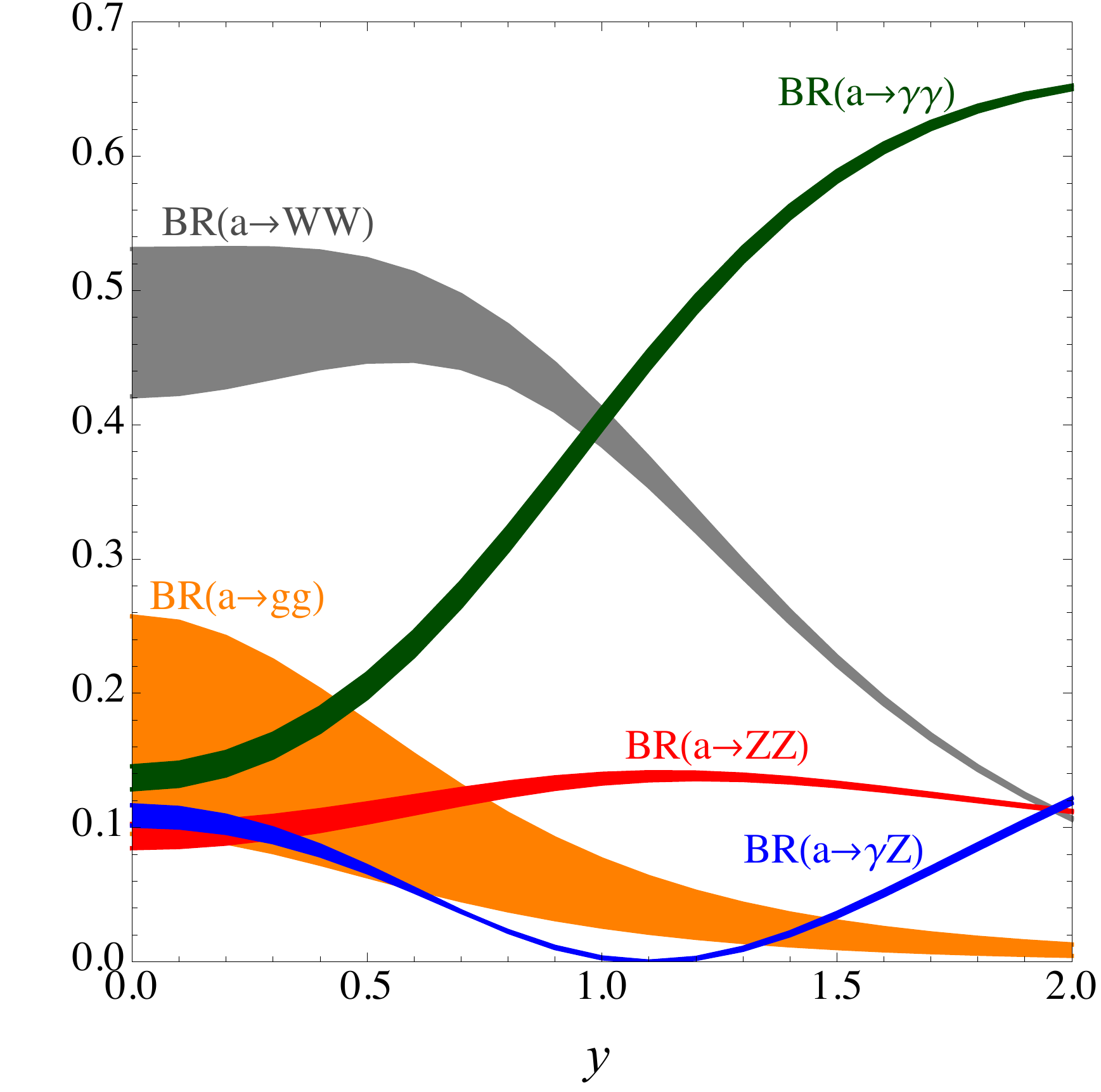}
\caption{{\it Left panel}: Correlation between $y_T$ and  the  hypercharge $y$ of the fundamental (left-handed) technifermions that allows to explain
the diphoton excess in the $\eta^{\prime}$-like scenario. Region 2 and region 3 correspond to large couplings of the $a$ to the vectorlike quark $T$. {\it Right panel}: 
Branching ratios for the decay of $a$ into a pair of gauge bosons predicted in the $\eta^\prime$-like scenario as a function of $y$, calculated for values of $y_T\geq 0$  in region 1. For $y_T=0$ the production of the resonance is entirely given by the photon fusion process.}
\label{fig:BR}
\end{center}
\end{figure*}

We show in the left panel of Fig.~\ref{fig:BR} 
$y_T$ versus  $y$ regions which reproduce  the diphoton excess within $\pm 1\sigma$.  The results do not change significantly by varying $d(R)$. The $(y_T,y)$ solutions, for example, that we obtain for $d(R)=4$ overlap with those for $d(R)=6$. 
We have three different regions of $(y_T,y)$ parameters for which the diphoton excess can be reproduced, as shown in the left panel of Fig.~\ref{fig:BR}. Region 1 corresponds to two solutions  with almost identical small values of $|y_T|$, in agreement with our naive expectation from EGD. Region 2 and 3 correspond to negative $y_T$ values and fall in the large-coupling regime of the $a$ coupling to $T$, $|y_T| m_T/F_T > \sqrt{4 \pi}$. 
We include in the calculation of the production cross-section the photon fusion mechanism \cite{Fichet:2013gsa}, which gives a contribution larger than 50\% (smaller than 10\%) to the total cross-section 
for $y\gtrsim 1.2$ ($y\lesssim 0.7$) \cite{wip}.

The predictions for the branching ratios of $a$ decaying into the possible different pairs of EW gauge bosons and gluons are reported
in the right panel of Fig.~\ref{fig:BR} and are calculated for the positive $y_T$ solutions in the perturbative region 1. 
The $\eta^\prime$-like explanation of the diphoton excess passes all of the current constraints from the LHC searches for diboson resonances. At $\sqrt{s}=13$ TeV, for a resonance of $\sim$750 GeV, the strongest constraint on the $ZZ$ 
($WW$) channel is placed by the ATLAS search in \cite{ATLAS-CONF-2015-071} (\cite{ATLAS-CONF-2015-075}) which gives an upper limit of about 250 (250) fb on $\sigma\left(pp \to a \to ZZ~(WW)\right)$, while for our $\eta^{\prime}$ we have at most $\sim$4 (20) fb. The run-1 searches at $\sqrt{s}=8$ TeV give an upper limit on the resonance cross-section of about 10 fb in the $ZZ$ channel \cite{Aad:2015kna}, to be compared with at most $\sim$1 fb for our $\eta^\prime$, and of about 40 fb in the $WW$ channel \cite{Aad:2015agg}, where we have at most $\sim$4 fb. 
Finally, the ATLAS search on the $Z\gamma$ channel \cite{Aad:2014fha} gives an upper limit $\sigma(pp \to a \to Z\gamma)\lesssim 4$ fb, which is fulfilled in our scenario, predicting a cross-section of 
at most $\sim$1 fb.

The di-jet channel from the $a \to gg$ decay, which is the dominant decay mode for the axion-like scenario discussed in the first part of this work is instead suppressed for the $\eta^\prime$-like state. The $WW$ channel, which is absent for the axion-like particle may have a relevant branching ratio which is enhanced compared to $\gamma\gamma$ for $y\lesssim 1$. The $ZZ$ channel is also enhanced compared to the axion-like scenario, especially for smaller $y$ values.
Finally, the $\gamma Z$ channel is enhanced compared to the axion-like case for $y\lesssim 0.5$, whereas it is suppressed for larger $y$ values.

In the present study we have not included the potential effects of a direct SM top coupling to either the elementary or composite $a$ particle. In the elementary case we have checked that,  when one includes the coupling to the top  $m_{t}/f_a\sim 0.7$,  the branching ratio in diphoton is too small to explain the excess because the total width of $a$ is dominated by the tree-level decay into SM tops. 
 In the composite case the actual strength of the coupling to the SM top depends on the underlying scenario for the SM fermion mass generation. To provide a direct comparison with the axion case  we have considered above scenarios in which the coupling to the SM top is suppressed. However,  we have also investigated the case in which this coupling is as big as $m_{t}/F_T$. We find that the excess can be reproduced, unlike the axion case, because of the topological terms, with and without the inclusion of the extra $T$ fermion. We also checked that this scenario is consistent with the run-1 LHC experimental bounds on  $\bar{t}t$ resonances \cite{CMS:2015nza, Aad:2015fna}.

Our results show that a highly natural and minimal composite nature of the new potential particle, in terms of an $\eta^\prime$-like state, decaying into two photons can explain the excess. We have also demonstrated that the underlying axion or minimal composite nature of this state can be disentangled upon discovery and careful analysis of the related decay channels.  Furthermore, 
our analysis is immediately applicable to set relevant  constraints on EW scale composite dynamics at run-2 LHC.

{\it Acknowledgments and note added:}  The CP$^3$-Origins center is partially funded by the Danish National Research Foundation, grant number DNRF90. While this paper was being completed a number of papers appeared \cite{Pilaftsis:2015ycr,Mambrini:2015wyu}  that partially overlapped with the present one.

\end{document}